\documentclass[12pt]{article}

\usepackage{amsmath,amssymb,amsthm,mathtools,bm}
\usepackage{setspace,xcolor,enumitem,url,booktabs,soul,multirow,arydshln,framed,hyperref}
\usepackage{tikz-cd}
\usepackage{blkarray}
\usepackage{natbib,natbibspacing}

\newtheorem{corollary}{Corollary}
\newtheorem{lemma}{Lemma}

\usepackage{authblk}
\author[1,2]{Melike Oguz-Alper\footnote{Address for correspondence: B58, West Highfield Campus, University of Southampton, SO17 1BJ, Southampton, United Kingdom}}
\author[2,1]{Li-Chun Zhang}
\affil[1]{\em \small University of Southampton, United Kingdom}
\affil[2]{\em \small Statistics Norway, Norway}
\title{Snowball sampling from graphs}
\date{}

\begin{document}

\maketitle

\begin{abstract} We develop unbiased strategies to probabilistic $T$-wave snowball sampling from graphs, where the interest of estimation may concern  finite-order subgraphs such as triangles, cycles or stars. Our approaches encompass also the finite-population sampling strategies to multiplicity sampling and adaptive cluster sampling, both of which can be recast as snowball sampling aimed at graph node totals. A general snowball sampling theory offers greater flexibility in terms of scope and efficiency of graph sampling, in addition to the existing random node or edge sampling methods. 
\end{abstract}


\section{Introduction} 

By \emph{snowball sampling (SBS)} starting from an initial sample of units, one would include any additional units that are outside the current sample but can be reached from them at each wave according to a specified \emph{observation procedure}, and the process is either terminated after a prefixed number of waves (denoted by $T$) or if no new units can be added. Thus, the sample is observed in the same way as by \emph{breadth-first search} in a graph, denoted by $G = (U,A)$, where the nodes $U$ represent all the units that can be observed by SBS following all possible initial samples, and the edges $A$ represent the links among the units to which the observation procedure is applied at each wave. 

SBS is not a probability sampling method, if the initial sample of nodes do not have known sampling probabilities, such as when they are obtained merely by convenience, or if the transition from one wave to the next does not strictly follow the specified observation procedure, such as when it relies on the cooperation of survey respondents who do not implement it faithfully. It may still be possible to estimate the parameter on interest by an assumed model of the sampling process, such as the case with respondent driven sampling \citep[e.g.][]{SalganikHeckathorn2004, Gileetal2015}.

We consider probabilistic SBS from graphs, where the initial node sample is selected with known probabilities and the subsequent observation procedure strictly follows its specification. \citet{Goodman1961} studies a special case of SBS from this perspective, where the out-degree (best friend) is one for every node in a directed graph, and the number of reciprocal edges (mutual best friends) is to be estimated. \citet{Frank1977c,Frank1979} and \citet{FrankSnijders1994} consider one-wave SBS from simple graphs to estimate totals over all the nodes. \citet{Rebecq2018} considers one-wave SBS and adaptive cluster sampling \citep{Thompson1990} following initial Bernoulli sampling to estimate the total number of nodes with specific characteristics in a large network such as Twitter.

Notice that \citet{Frank1971,Frank1977a,Frank1978a,Frank1978b,Frank1980a,Frank1981} considers estimation for specific subgraphs (such as 2-star or triangle) by graph sampling based on the \emph{induced} observation procedure, where an edge is observed only if both its end nodes are included in the initial sample, and the probability of observing the subgraphs derives directly from the initial sampling design since no additional sample nodes can be added to the initial sample by induced observation. In contrast, the observation procedure of SBS from graphs is \emph{incident}, where an edge may be observed from one of its end nodes in the current sample. While this makes it possible to include additional sample nodes wave-by-wave, it also complicates the calculation of the associated sampling probabilities, because a subgraph can be observed in many ways and not all of them need to be realised on a given sampling occasion. Indeed, \citet{ZhangPatone2017} derive the probability of observing any given subgraph under $T$-wave SBS, $T\geq 1$, without explaining how the probability can be obtained when the whole graph is only partially observed. 

\begin{figure}[ht]
\centering
\begin{tikzcd}
\cdots g_1 \arrow[r, dash] & h_1 \arrow[r, dash] & j_3 \arrow[d, dash] \arrow[dr, dash] \arrow[r, dash] & i_3 \arrow[dr, dash] \arrow[d, dash] 
& h_2 \arrow[r, dash] & g_2 \cdots & \\    
\cdots h_3 \arrow[r, dash] & j_1 \arrow[r, dash] & i_1 \arrow[r, dash] \arrow[ur, dash] & i_2 \arrow[r, dash] & j_2 \arrow[u, dash] \arrow[r, dash] & h_4 \cdots
\end{tikzcd} 
\caption{A part of the population graph  in Figure \ref{fig:popgraph}.} \label{fig:TSBS}
\end{figure}

Thus, a general theory for $T$-wave SBS from graphs is needed, where the estimation may concern finite-order subgraphs not limited to the nodes. For example, consider the node set $M = \{ i_1, i_2, i_3\}$ in an undirected simple graph illustrated in Figure \ref{fig:TSBS}. The specific characteristics of the subgraph $G(M)$ induced by $M$ will be referred to as \emph{motif}, which is triangle in this case, and the \emph{order} of a motif is simply that of $G(M)$, which is $|M|$. A motif is said to be \emph{induced} \citep{Zhang2021} if it can be determined from the corresponding subgraph $G(M)$ alone, regardless how this subgraph may be connected to the rest of the graph. For instance, triangle is an induced motif, but component is not. 

\begin{figure}[ht] 
\centering
\begin{tikzcd}
& & \circ \arrow[dr, dash] & & \circ \arrow[d, dash] & & \circ  \arrow[d, dash] \arrow[dr, dash] &  \\
& \circ & & \circ & \circ \arrow[r, dash] & \circ & \circ \arrow[r, dash] & \circ 
\end{tikzcd}
\[
\hspace{3mm} \mathcal{K}_1 \hspace{15mm} \mathcal{K}_2  \hspace{25mm} \mathcal{S}_2  \hspace{25mm} \mathcal{K}_3
\]
\begin{tikzcd}
& \circ \arrow[d, dash] \arrow[r, dash] & \circ & \circ \arrow[d, dash] \arrow[r, dash] & \circ \arrow[dl, dash] 
& \circ & \circ & \circ \arrow[d, dash] & \circ \\
& \circ & \circ \arrow[l, dash] \arrow[u, dash] & \circ & \circ \arrow[l, dash] \arrow[u, dash] \arrow[ul, dash] 
& \circ \arrow[u, dash] \arrow[ur, dash] \arrow[r, dash] & \circ & \circ & \circ \arrow[l, dash] \arrow[u, dash]
\end{tikzcd}
\[
\hspace{11mm} \mathcal{C}_4 \hspace{22mm} \mathcal{K}_4 \hspace{23mm} \mathcal{S}_3 \hspace{23mm} \mathcal{P}_3
\]
\caption{Node ($\mathcal{K}_1$), 2-clique (edge, $\mathcal{K}_2$), 2-star ($\mathcal{S}_2$), 3-clique (triangle, $\mathcal{K}_3$), 4-cycle ($\mathcal{C}_4$), 4-clique ($\mathcal{K}_4$), 3-star ($\mathcal{S}_3$) and 3-path ($\mathcal{P}_3$) as induced motifs in undirected simple graphs.} \label{fig:motifs}
\end{figure}

In this paper we consider induced motifs of connected subgraphs in simple graphs as the entities of estimation. Some examples of such low-order induced motifs are given in Figure \ref{fig:motifs}. Let $\Omega$ contain all the occurrences of some particular motifs in a given \emph{population graph} $G$, which are of interest, such as all the triangles in $G$. For each $\kappa \in \Omega$, defined for the corresponding subgraph $G(M_{\kappa})$ induced by the nodes $M_{\kappa}$, let $y_{\kappa}$ be a fixed value associated with $\kappa$. The corresponding \emph{graph total} over $\Omega$ is given by
\begin{equation} \label{theta}
\theta = \sum_{\kappa \in \Omega} y_{\kappa} ~.
\end{equation}

In Section \ref{sec:theory} we shall develop design-based unbiased \emph{strategies} to the estimation of $\theta$ under SBS from graphs, where a strategy consists of a method of sampling and an associated estimator, denoted by $\hat{\theta}$, and a strategy is unbiased if 
\[
E(\hat{\theta}) = \theta
\]
where the expectation is taken over hypothetically repeated SBS from the given $G$.

As will be explained in Section \ref{sec:node}, our approaches encompass also the existing finite-population sampling strategies to \emph{multiplicity sampling} \citep[e.g.][]{BirnbaumSirken1965,Sirken1970,Sirken2005,Lavallee2007} and \emph{adaptive cluster sampling} \citep{Thompson1990, Thompson1991}. This is because both these sampling methods can be recast as SBS, where all the units that can be reached at a given wave (according to the specified observation procedure) are included in the sample, although the parameter of interest for these finite-population problems has traditionally been limited to the order-1 motif, i.e. node. 

Estimation of graph totals of other finite-order motifs will be illustrated and discussed in Section \ref{sec:motif}. Some final remarks and topics for future research are given in Section \ref{sec:final}.

\section{Theory} \label{sec:theory}

Given a (population) graph $G = (U, A)$, let an initial node sample $s_0$ be selected from $U$ according to a known sampling design. Let $\alpha(s_0)$ consist of the nodes that can be observed from $s_0$ next. Specifically, for simplicity of exposition, we assume \emph{reciprocal incident observation procedure (RIOP)} throughout this section, whereby any node $j$ adjacent to a node $i$ in $s_0$ is observed, as well as the edge(s) between $i$ and $j$, whether the graph is directed or undirected. For $t=1,2,\dots, T$, let the \emph{$t$-th wave} node sample be
\begin{eqnarray*}
s_t=\alpha(s_{t-1})\setminus \bigcup_{r=0}^{t-1}s_r ~.
\end{eqnarray*}
The \emph{$T$-wave SBS ($T$SBS)} is terminated once $s_T$ is obtained; if $s_t = \emptyset$ for some $t< T$, simply let $s_r = \emptyset$ for $r = t, ..., T$. The \emph{seed sample} of $T$SBS, defined as the nodes to which the specified OP has been applied \citep{Zhang2021}, is given by
\begin{eqnarray}
\label{eq:sets}
s= \bigcup_{t=0}^{T-1}s_t  ~.
\end{eqnarray}
The \emph{sample graph} by $T$SBS is the observed subgraph of $G$, which is given by 
\[
G_s=(U_s,A_s)\quad\text{and}\quad A_s= A\cap s_{ref} \quad\text{and}\quad U_s=s\cup \mbox{Inc}(A_s)
\]
where $s_{ref} = s\times U \cup U \times s$ refers to the observed part of the adjacency matrix of $G$ according to the RIOP, and $\mbox{Inc}(A_s)$ contains the nodes that are incident to $A_s$. 

Denote by $\Omega_s$ the observed elements of $\Omega$ in $G_s$. Any node $i$ is called a \emph{($T$SBS) ancestor} of $\kappa$, $\forall \kappa\in \Omega$, if the selection of $i$ in $s_0$ alone would imply the observation of $\kappa$ in $G_s$, i.e.
\[
\Pr(\kappa \in \Omega_s \mid s_0 = \{ i \}) = 1 ~.
\]
Denote by $\beta_{\kappa}$ all the ancestors of $\kappa$ in $G$. Let $\beta_{\kappa}^*$ be a chosen, \emph{fixed} subset of ancestors, where $\beta_{\kappa}^* \subseteq \beta_{\kappa}$. Let the corresponding sampling probability 
\begin{equation} \label{eq:pik}
\pi_{(\kappa)} = \mbox{Pr}(\beta_{\kappa}^* \cap s_0 \neq \emptyset) 
\end{equation}
be the probability that at least one node in $\beta_{\kappa}^*$ is selected initially. The estimator 
\begin{equation} \label{eq:that}
\hat{\theta} = \sum_{\kappa\in \Omega} \mathbb{I}(\beta_{\kappa}^* \cap s_0 \neq \emptyset) \frac{y_{\kappa}}{\pi_{(\kappa)}}
\end{equation}
is unbiased for $\theta$ under $T$SBS, where $\mathbb{I}(\cdot)$ is the 1/0-indicator function. Moreover, let  
\[
\pi_{(\kappa \ell)} = \mbox{Pr}(\beta_{\kappa}^* \cap s_0 \neq \emptyset,~ \beta_{\ell}^* \cap s_0 \neq \emptyset) 
\]
be the joint sampling probability of $\kappa, \ell\in \Omega$. The SBS variance of $\hat{\theta}$ is given by
\[
V(\hat{\theta}) = \sum_{\kappa, \ell\in \Omega} \Big( \frac{\pi_{(\kappa \ell)}}{\pi_{(\kappa)} \pi_{(\ell)}} - 1\Big) y_{\kappa} y_{\ell} ~.
\]

Note that \eqref{eq:pik} and \eqref{eq:that} can be calculated only if, for any $\kappa \in \Omega$, one is able to determine $\mathbb{I}(\beta_{\kappa}^* \cap s_0 \neq \emptyset)$ given any possible sample graph $G_s$ where $\kappa$ is observed, provided which any choice of $\beta_{\kappa}^*$ would yield an unbiased strategy for $\theta$ accordingly. Below we first outline three possible strategies to the choice of $\beta_{\kappa}^*$, illustrated for the triangle $\kappa$ with nodes $M_{\kappa} = \{ i_1, i_2, i_3\}$ in Figure \ref{fig:TSBS}, before moving on to more detailed development.

\paragraph{I} Let $\beta_{\kappa}^* = \beta_{\kappa}$ for any $\kappa \in \Omega$, consisting of all its $T$SBS ancestors. 

Note that one cannot always observe $\beta_{\kappa}$ entirely in a given $G_s$. For instance, under 3SBS starting from $s_0 = \{ h_1\}$ in Figure \ref{fig:TSBS}, one would not observe that $h_2$ is also a 3SBS-ancestor of $\kappa$. Thus, additional observations to $G_s$ are generally needed to implement this strategy. Although this is unattractive in many situations, it may still be worth considering, when SBS is carried out to provide a compressed view of a known but very large graph, in order to avoid processing the whole graph. 

\paragraph{II} Let $\beta_{\kappa}^* \subseteq \beta_{\kappa}$ be a \emph{restricted} set of ancestors for any $\kappa \in \Omega$, provided the \emph{same} $\beta_{\kappa}^*$ can be identified \emph{whenever} $\beta_{\kappa}^* \cap s_0 \neq \emptyset$, i.e. whenever $\kappa$ is observed in the sample graph starting from any node in $\beta_{\kappa}^*$.  

For instance, under 3SBS given $|s_0|=1$ in Figure \ref{fig:TSBS}, $\beta_{\kappa}^* = \{ i_1, i_2, i_3, j_1, j_2, j_3 \}$ can always be identified as 3SBS ancestors whenever $\beta_{\kappa}^* \cap s_0 \neq \emptyset$. Unlike Strategy-I, this one does not require additional observations to $G_s$.

\paragraph{III} Let $\beta_{\kappa}^* \subseteq \beta_{\kappa|s}$ for any $\kappa \in \Omega$, where $\beta_{\kappa|s}$ denotes all its $T$SBS ancestors that can be verified in the realised sample graph $G_s$. 

Unlike Strategy-I or II, this one is said to be \emph{sample dependent} in the sense that another $\beta_{\kappa}^*$ could have been chosen insofar as $\beta_{\kappa|s}$ is unknown in advance. For example, one can let $\beta_{\kappa}^* \subseteq \beta_{\kappa|s} = \{ i_1, i_2, i_3, j_1, j_2, j_3, h_1 \}$ if 3SBS starts from $s_0 =\{ h_1\}$ in Figure \ref{fig:TSBS}; whereas $h_1 \in \beta_{\kappa}^*$ would not have been possible had 3SBS started from $s_0 = \{ h_2\}$. This strategy does not require additional observation to $G_s$ either, and it is computationally the least demanding of all the strategies outlined here.

\subsection{Strategy-I} \label{sec:I}

The induced (motif of) subgraph $G(M_{\kappa})$ is observed iff $M_{\kappa}\times M_{\kappa}\subseteq s_{ref}$. The \emph{observation distance} from a node $i$ to $\kappa$ is the number of waves it takes from $i$ to observe $G(M_{\kappa})$, denoted by $d_{i,\kappa}$. A node $i$ is a $T$SBS ancestor of $\kappa$ if $d_{i,\kappa} \leq T$. Below we first explain the calculation of $d_{i,\kappa}$ and then provide the result that is needed to use $\beta_{\kappa}^* = \beta_{\kappa}$ in \eqref{eq:pik}. 

Let $\varphi_{ij}$ be the \emph{geodesic distance} from node $i$ to $j$ in $G$, which is the number of waves it takes from $i$ to observe $j$. For example, in Figure \ref{fig:TSBS}, we have $\varphi_{i_1 i_2}=\varphi_{j_3 i_2}=1$, $\varphi_{j_3 j_2}=\varphi_{h_1 i_2}=2$, and $\varphi_{g_1 g_2}=\varphi_{h_3 g_2}=6$. Let $\varphi_{ij} =\infty$ if $j$ cannot be observed from $i$. Given the RIOP, $\varphi_{ij} = \varphi_{ji}$ whether the graph is directed or undirected. 

The \emph{geodesic distance} from a node $i$ to a motif $\kappa$ in $G$ is defined as the number of waves it takes from $i$ to observe \emph{any} of the nodes $M_{\kappa}$, and it is given by
\begin{eqnarray}
\label{eq:geodistik}
\varphi_{i,\kappa}= \begin{cases}
\min_{j\in M_{\kappa}}\varphi_{ij} & \text{if } i\notin M_{\kappa} \\
0 & \text{if } i\in M_{\kappa} \end{cases} ~.
\end{eqnarray}
For the motif $\kappa$ with $M_{\kappa}=\{i_1,i_2,i_3\}$ in Figure \ref{fig:TSBS}, we have $\varphi_{i_1,\kappa}=\varphi_{i_2,\kappa}=\varphi_{i_3,\kappa}=0$, $\varphi_{j_1,\kappa}=\varphi_{j_2,\kappa}=\varphi_{j_3,\kappa}=1$, $\varphi_{h_1,\kappa}=\varphi_{h_2,\kappa}=\varphi_{h_3,\kappa}=\varphi_{h_4,\kappa}=2$ and $\varphi_{g_1,\kappa}=\varphi_{g_2,\kappa}=3$.

The number of waves it takes from a node $i$ to observe \emph{all} the nodes $M_{\kappa}$ is called the \emph{radius distance} from $i$ to $\kappa$ in $G$, and it is given by
\begin{eqnarray*}
\label{eq:radiusdistik}
\lambda_{i,\kappa}=
\max_{j\in M_{\kappa}}\varphi_{ij} ~.
\end{eqnarray*}
For the motif $\kappa$ with $M_{\kappa}=\{i_1,i_2,i_3\}$ in Figure \ref{fig:TSBS}, we have $\lambda_{i_1,\kappa}=\lambda_{i_2,\kappa}=\lambda_{i_3,\kappa}=\lambda_{j_3,\kappa}=1$, $\lambda_{j_1,\kappa}=\lambda_{j_2,\kappa}=\lambda_{h_1,\kappa}=2$, $\lambda_{h_2,\kappa}=\lambda_{h_3,\kappa}=\lambda_{h_4,\kappa}=\lambda_{g_1,\kappa}=3$ and $\lambda_{g_2,\kappa}=4$.

\begin{lemma} \label{lemma:dik}
For a node $i$ and an induced motif $\kappa$ of connected $G(M_{\kappa})$, we have 
\[
d_{i,\kappa} = \begin{cases} \lambda_{i,\kappa} & \text{if } | \arg \max_{j\in M_{\kappa}} \varphi_{ij} | = 1 \\
1 + \lambda_{i,\kappa} & \text{otherwise} \end{cases} ~.
\]
\end{lemma}

\begin{lemma} \label{lemma:ancestralT}
For any induced motif $\kappa$ of connected $G(M_{\kappa})$, which is observed by $T$SBS, one needs at most $T - \mathbb{I}(|M_{\kappa}| >1)$ additional waves to observe $\beta_{\kappa}$. 
\end{lemma}

The Lemmas \ref{lemma:dik} and \ref{lemma:ancestralT} can be applied for Strategy-I; see Appendix \ref{proofs} for the proofs. 
Notice that there is a possibility of empty ancestor set $\beta_{\kappa} = \emptyset$ if $T$ is too small, in which case one can increase $T$ till $d_{i,\kappa} \leq T$ for at least some $i\in M_{\kappa}$. For example, for any triangle $\kappa$, we have $\beta_{\kappa} = \emptyset$ under 1SBS and $M_{\kappa} \subseteq \beta_{\kappa}$ under 2SBS, because a triangle can only be observed in two waves from any of its nodes $M_{\kappa}$. Meanwhile, it is still possible to observe the triangle $\kappa$ by 1SBS if at least two nodes in $M_{\kappa}$ are selected in $s_0$. Similarly for many other motifs. Thus, more generally for $T$SBS, one can allow sets of nodes as the \emph{joint} ancestors of a given $\kappa$, and let all its $T$SBS ancestors be given by 
\[
\mathcal{B}_{\kappa} = \{ M \subset U : d_{M,\kappa} \leq T,~ |M| \leq |s_0| \}
\]
where $d_{M,\kappa}$ is the observation distance from a node set $M$ to a motif $\kappa$. An unbiased estimator $\hat{\theta}$ can then be given similarly to \eqref{eq:that} on replacing $\mathcal{B}_{\kappa}^*$ for $\beta_{\kappa}^*$. However, we do not pursue such a generalisation in this paper because we have not yet clarified all the computational aspects.

\subsection{Strategy-II}  \label{sec:II}

By this strategy one must be able to verify the \emph{same} restricted ancestor set $\beta_{\kappa}^*$, whenever $\kappa$ is observed in $G_s$ starting from any node in $\beta_{\kappa}^*$. Below we provide some results by which $\beta_{\kappa}^*$ can be specified regardless the realised $G_s$. The proofs are given in  Appendix \ref{proofs}.

\begin{corollary} \label{cor:TforbetaK}
For any induced motif $\kappa$ of connected $G(M_{\kappa})$, a restricted set of $T$SBS ancestors is given by
\[
\beta_{\kappa}^* = \{ i\in U : d_{i,\kappa} \leq t_{\kappa} \} \quad\text{and}\quad
t_{\kappa} = \lfloor \{ T+ \mathbb{I}([M_{\kappa}| >1) \} /2 \rfloor ~.
\]
\end{corollary}

Corollary \ref{cor:TforbetaK} follows directly from Lemma  \ref{lemma:ancestralT}. It can be quite restrictive. One can often obtain a larger restricted set by the next result, for which it is convenient to introduce 
\begin{eqnarray*}
\label{eq:diam}
\varphi_{\kappa}=\max_{i,j\in M_{\kappa}} \varphi_{ij}
\end{eqnarray*}
as the \emph{diameter} of a given motif $\kappa$ in $G$, and
\begin{eqnarray*}
\zeta_{\kappa}=\max_{i\in M_{\kappa}}~ d_{i,\kappa}
\end{eqnarray*}
as its \emph{observation diameter} in $G$. By Lemma \ref{lemma:dik}, if $\varphi_{\kappa}<\infty$, then we have
\begin{eqnarray*}
\zeta_{\kappa}\leq 1+\varphi_{\kappa} ~.
\end{eqnarray*}

\begin{lemma} \label{lemma:combineddefault}
For any induced motif $\kappa$ of connected $G(M_{\kappa})$, a restricted set of $T$SBS ancestors is given by
\[
\beta_{\kappa}^* = \{ i\in U :d_{i,\kappa} \leq t_{\kappa} \}\cup \{ i\in M_{\kappa} : d_{i,\kappa} \leq T\} 
\cup \{ i\not \in M_{\kappa} : \varphi_{i,\kappa} \leq t_{geo} \}
\]
where $t_{\kappa}$ is as given in Corollary \ref{cor:TforbetaK}, and $t_{geo} = \max\big( 0,  \lfloor (T - \varphi_{\kappa})/2 \rfloor\big)$.
\end{lemma}

The restricted set given by Corollary \ref{cor:TforbetaK} is clearly a subset of that by Lemma \ref{lemma:combineddefault}. But there is still room for improving Lemma \ref{lemma:combineddefault}. Take the triangle with $M_{\kappa} = \{ i_1, i_2, i_3 \}$ in Figure \ref{fig:TSBS}. Corollary \ref{cor:TforbetaK} is not effective under 2SBS, because $t_{\kappa} = 1$ but $d_{i,\kappa} \geq 2$ for any node $i$. We have $t_{geo} =0$ for Lemma \ref{lemma:combineddefault}, so that $\beta_{\kappa}^* = M_{\kappa}$ because $\{ i\not \in M_{\kappa} : \varphi_{i,\kappa} \leq t_{geo} \}$ is empty. The result below can be useful when $T$ is even.  

\begin{lemma} \label{lemma:simpleextension}
For any induced motif $\kappa$ of connected $G(M_{\kappa})$, where $|M_{\kappa}| >1$, let $\beta_{\kappa}^{\bullet}$ be given by Lemma \ref{lemma:combineddefault} for $T$SBS, where $T$ is even. A restricted set of $T$SBS ancestors is given by
\[
\beta_{\kappa}^* = \beta_{\kappa}^{\bullet} \cup \{ i\notin \beta_{\kappa}^{\bullet}:\lambda_{i,\kappa}\leq T/2 \}
\]
\end{lemma}

By Lemma \ref{lemma:simpleextension} one can now include $j_3$ in $\beta_{\kappa}^*$ for 2SBS in addition to $M_{\kappa} = \{ i_1, i_2, i_3\}$. Meanwhile, we have $\beta_{\kappa} = M_{\kappa} \cup \{ j_2,j_3\}$ under 2SBS in this graph, and all the nodes $\beta_{\kappa}$ can be identified as 2SBS ancestors whenever $\kappa$ is observed, where $\kappa$ can be observed from either $j_2$ or $j_3$ as fast as from any node in $M_{\kappa}$. It may therefore still be possible to improve the Lemmas \ref{lemma:combineddefault} and \ref{lemma:simpleextension} by future research.

\subsection{Strategy-III} \label{sec:III}

Recall that by definition $\beta_{\kappa|s}$ contains all the $T$SBS ancestors, for any given $\kappa \in\Omega_s$, which can be identified in the realised sample graph $G_s$. Let $d_{i,\kappa}(G_s)$ be the \emph{sample observation distance} calculated in $G_s$ (as if it were $G$), where $d_{i,\kappa}(G_s) \geq d_{i,\kappa}(G)$ generally.

\begin{lemma} \label{lemma:betas}
For any induced motif $\kappa$ of connected $G(M_{\kappa})$ observed by $T$SBS, we have 
\[
\beta_{\kappa|s} = \{  i \in U_s : d_{i,\kappa}(G_s) \leq T \} ~.
\]  
\end{lemma} 

The Lemma \ref{lemma:betas} establishes $\beta_{\kappa|s}$ as a viable strategy; see Appendix \ref{proofs} for the proof.
Note that for each $i\in \beta_{\kappa|s}$, there exists at least a path corresponding to $d_{i,\kappa}(G_s)$ in the sample subgraph that is induced by $\beta_{\kappa|s}$, denoted by $G_s\big(\beta_{\kappa|s}\big)$, whose edge set is given by $\{ (ij) \in A_s : i,j \in \beta_{\kappa|s} \}$ instead of $\{ (ij) \in A : i,j \in \beta_{\kappa|s} \}$. This is because any node on the path giving rise to $d_{i,\kappa}(G_s)$ must have a shorter sample observation distance to $\kappa$ than $i$, which means that it must belong to $\beta_{\kappa|s}$. Thus, to identify $\beta_{\kappa|s}$, one simply needs to check, for $t=1, 2, ...$, the sample subgraph $G_s(\alpha_{\kappa}^t)$ induced by the nodes
\[
\alpha_{\kappa}^t = \alpha(\alpha_{\kappa}^{t-1})
\]
where $\alpha_{\kappa}^0 = M_{\kappa}$ by default, to find any node $i$ in $\alpha^t \setminus \alpha_{\kappa}^{t-1}$ with $d_{i, \kappa}\big(G_s(\alpha_{\kappa}^t) \big) \leq T$, until no such nodes exist at some $t$.  

Take $M_{\kappa} = \{ i_1, i_2, i_3 \}$ in Figure \ref{fig:TSBS} for an illustration. We have $\beta_{\kappa|s} = \beta_{\kappa} = \emptyset$ for 1SBS, and $\beta_{\kappa |s} = \beta_{\kappa} = M_{\kappa} \cup \{ j_2, j_3 \}$ for 2SBS. For 3SBS, $\beta_{\kappa} = M_{\kappa} \cup \{ j_1, j_2, j_3 \} \cup \{ h_1, h_2, h_4\}$ and it is possible to let
\[
\beta_{\kappa |s} = \begin{cases} M_{\kappa} \cup \{ j_2, j_3 \} \cup \{ j_1, h_1\} & \text{if } s_0 \cap \beta_{\kappa} \subseteq \{ j_1, h_1\} \\
M_{\kappa} \cup \{ j_2, j_3 \} \cup \{ h_2, h_4\} & \text{if } s_0 \cap \beta_{\kappa} \subseteq \{ h_2, h_4\} \\
\beta_{\kappa} &  \text{if } s_0 \cap \beta_{\kappa} \neq \emptyset \text{ otherwise} \color{black} \end{cases} ~.
\]

Now, for any $G(M_{\kappa})$ with $|\beta_{\kappa}| >1$, it is possible for $\beta_{\kappa|s}$ to vary with the realised sample graphs $G_s$. One can envisage a sequence of sample graphs generated by repeated sampling from $G$, \color{black} by which $\beta_{\kappa|s}$ is fixed for every $\kappa$ (in $\Omega$) according to the first $G_s$ where $\beta_{\kappa|s} \cap s_0 \neq \emptyset$, i.e. the first time the motif $\kappa$ can be used for estimation by Strategy-III. Denote the resulting \emph{collection} by
\[
\beta_{\Omega|s} = \{ \beta_{\kappa|s} : \kappa\in \Omega\} ~.
\]
There is only a finite number of distinct collections $\beta_{\Omega|s}$, and each $\beta_{\Omega|s}$ yields a $T$SBS variance of the estimator \eqref{eq:that} under the Strategy-III. An illustration will be given in Section \ref{sec:multiplicity} later. Meanwhile, based on the sample graph $G_s$, an estimator of $V(\hat{\theta})$ requires only $\beta_{\kappa|s}$ for each $\kappa$ in $\Omega_s$. It follows that the variance estimator is aimed at the weighted average of the sampling variances corresponding to all the collections $\beta_{\Omega|s}$ that are compatible with the realised $\{ \beta_{\kappa|s} : \kappa \in \Omega_s\}$, where the weight of each $\beta_{\Omega|s}$ derives from the probability of its corresponding sequences of sample graphs starting with the realised $G_s$. Thus, a variance estimator that is unbiased for the conditional variance given $\beta_{\Omega|s}$ is also an unbiased estimator of the unconditional variance of Strategy-III over repeated sampling.

\section{SBS for population totals} \label{sec:node}

\subsection{Multiplicity sampling} \label{sec:multiplicity}

In finite-population multiplicity sampling, each measurement (or study) unit is accessed indirectly via one or several sampling units, such as when patients of a certain disease (measurement units) are sampled via the medical centres (sampling units) at which they receive treatments, 
or when a group of adult siblings (measurement units) are sampled via their respective households (sampling units) by referring to each. The multiplicity of a measurement unit is then the (number of) sampling units that can lead to it \citep{BirnbaumSirken1965}. Traditionally, the knowledge of multiplicity must be collected in the survey, in order to calculate the sampling probability of each observed measurement unit. 
Below we explain how one can recast finite-population multiplicity sampling as SBS from graphs, whereby unbiased estimation of $\theta$ defined by \eqref{theta} over all the measurement units becomes achievable even when the full knowledge of multiplicity cannot be obtained by surveying the sampled measurement units. 

Let the population graph be a bipartite graph $G = (F, \Omega; A)$, where $F$ denotes of all the sampling units and $\Omega$ all the measurement units, and $A$ denotes the edges that only exist from $F$ to $A$ but not between any two nodes in $F$ (or $\Omega$). Let $(i\kappa)\in A$ iff the selection of $i$ in $F$ leads to the observation of $\kappa$ in $\Omega$ according to the given multiplicity sampling method. For SBS from $G$, let the initial sample $s_0$ be selected from $F$, which leads to the sample of measurement units $\Omega_s$ by the RIOP, $\Omega_s \subseteq \Omega$. To collect the knowledge of multiplicity via survey is then the same as applying an extra 2nd wave of SBS from $\Omega_s$, yielding the 1SBS ancestor set $\beta_{\kappa} = \{ i\in F : (i\kappa)\in A\}$ for each sampled $\kappa$. In other words, traditional finite-population multiplicity sampling amounts to applying Strategy-I to 1SBS from such a bipartite graph.

\begin{figure}[ht]
\centering
\begin{tikzcd}
\kappa_1 & \kappa_2 & \kappa_3 & \kappa_4 & \kappa_5 & \kappa_6 & \kappa_7 \\
& i_1 \arrow[ul] \arrow[u] \arrow[ur] & i_2 \arrow[ul] \arrow[u] \arrow[urr] & i_3 \arrow[u] & i_4 \arrow[ull] \arrow[ul] \arrow[ur] \arrow[urr] \\
\kappa_1 & \kappa_2 & \kappa_3 & \kappa_4 & \kappa_5 & \kappa_6 & \kappa_7 \\
& i_1 \arrow[ul] \arrow[u] \arrow[ur] & i_2 \arrow[urr] & i_3 & i_4 \arrow[ull] \arrow[ul] \arrow[ur] \arrow[urr] \\
\kappa_1 & \kappa_2 & \kappa_3 & \kappa_4 & \kappa_5 & \kappa_6 & \kappa_7 \\
& i_1 \arrow[ul] & i_2 \arrow[ul] \arrow[u] \arrow[urr] & i_3 & i_4 \arrow[ull] \arrow[ul] \arrow[ur] \arrow[urr] 
\end{tikzcd} 
\caption{Illustration of multiplicity sampling as 1SBS from bipartite $G = (F, \Omega; A)$, top; two collections of sample-dependent $\beta_{\Omega|s}$, middle and bottom.} 
\label{fig:multiplicity}
\end{figure}

Suppose now the extra 2nd wave of SBS is infeasible due to the survey reality, such that the traditional methods \citep[e.g.][]{BirnbaumSirken1965,Sirken1970,Sirken2005,Lavallee2007} cannot be applied. Neither is Strategy-II feasible here for 1SBS by the Lemmas \ref{lemma:combineddefault} and \ref{lemma:simpleextension}. To illustrate how the sample-dependent Strategy-III can be applied for 1SBS in such situations, consider the simple bipartite graph in Figure \ref{fig:multiplicity} (top), with the sampling units $F = \{ i_1, ..., i_4\}$ and measurement units $\Omega = \{ \kappa_1, ..., \kappa_7\}$. Let $\theta = |\Omega| = 7$. Suppose simple random sampling without replacement (SRS) of $s_0$, $s_0\subset F$ and $|s_0| = 2$. 

As explained in Section \ref{sec:III}, let each collection of sample-dependent ancestor sets $\beta_{\Omega|s}$ be generated by a sequence of 1SBS from this graph. For instance, $\beta_{\Omega|s}$ depicted in the middle of Figure \ref{fig:multiplicity} can result from $s_0 = \{ i_1, i_4\}$ on the first occasion and $s_0 = \{ i_2, i_3\}$ on the second occasion. Notice that $\beta_{\kappa_2|s} = \{ i_1 \}$ first chosen given $s_0 = \{ i_1, i_4\}$ would remain fixed, although $i_2$ will be observed to be another ancestor of $\kappa_2$ later. In comparison, $\beta_{\Omega|s}$ depicted at the bottom of Figure \ref{fig:multiplicity} can result from $s_0 = \{ i_2, i_4\}$ on the first occasion and $s_0 = \{ i_1, i_3\}$ on the second occasion.

\begin{table}[ht]
\centering
\caption{Strategies of 1SBS for estimating $\theta = |\Omega|$. Strategy-I requires an extra wave of SBS. Multiplicity: size of $\beta_{\kappa|s}$ for $\kappa$ in $\Omega$, in italics if it can vary for Strategy-III.} 
\renewcommand{\tabcolsep}{0.20cm}
\renewcommand{\arraystretch}{1.2}
\begin{tabular}{lclcc} \toprule
Strategy & $\beta_{\kappa|s}=\beta_\kappa$ (if unstated) & Multiplicity & $\mathrm{CV}(|\Omega_s|)$ & $V(\hat{\theta})$   \\ \hline 
I &  & (1,2,3,2,1,1,1) & 0.202 &	2.987 \\ \hline
III, case 1 & $\beta_{\kappa_3|s}=\{1,2\}$, $\beta_{\kappa_4|s}=\{4\}$ & (1,\emph{2},\emph{2},\emph{1},1,1,1) &	0.322 & 6.333 \\ \hline
III, case 2 & $\beta_{\kappa_3|s}=\{1,2\}$, $\beta_{\kappa_4|s}=\{3\}$ & (1,\emph{2},\emph{2},\emph{1},1,1,1)	& 0.165	& 1.000 \\ \hline
III, case 3 & $\beta_{\kappa_3|s}=\{1,2\}$ & (1,\emph{2},\emph{2},\emph{2},1,1,1) & 0.248 & 3.507 \\ \hline
\multirow{2}{*}{III, case 4} & $\beta_{\kappa_2|s}=\beta_{\kappa_3|s}=\{1\}$, $\beta_{\kappa_4|s}=\{3\}$ & \multirow{2}{*}{(1,\emph{1},\emph{1},\emph{1},1,1,1)} & \multirow{2}{*}{0.274} & \multirow{2}{*}{3.667} \\ \cline{2-2}
& $\beta_{\kappa_2|s}=\beta_{\kappa_3|s}=\{2\}$, $\beta_{\kappa_4|s}=\{3\}$ &  &	 & \\ \hline
\multirow{2}{*}{III, case 5} & $\beta_{\kappa_2|s}=\{1\}$, $\beta_{\kappa_3|s}=\{1,4\}$, $\beta_{\kappa_4|s}=\{4\}$ & \multirow{2}{*}{(1,\emph{1},\emph{2},\emph{1},1,1,1)} & \multirow{2}{*}{0.410} & \multirow{2}{*}{8.467} \\ \cline{2-2}
& $\beta_{\kappa_2|s}=\{2\}$, $\beta_{\kappa_3|s}=\{2,4\}$, $\beta_{\kappa_4|s}=\{4\}$ &  &	 & \\ \hline
III, case 6 &$\beta_{\kappa_3|s}=\{4\}$ & (1,\emph{2},\emph{1},\emph{2},1,1,1) &	0.322 & 6.653 \\ \hline
\multirow{2}{*}{III, case 7} & $\beta_{\kappa_2|s}=\{1\}$, $\beta_{\kappa_3|s}=\{4\}$ & \multirow{2}{*}{(1,\emph{1},\emph{1},\emph{2},1,1,1)} & \multirow{2}{*}{0.351} & \multirow{2}{*}{6.867} \\ \cline{2-2}
& $\beta_{\kappa_2|s}=\{2\}$, $\beta_{\kappa_3|s}=\{4\}$  &  &	 & \\ \bottomrule
\end{tabular} \label{table:multiplicity} \end{table}

For the population graph at the top of Figure \ref{fig:multiplicity}, Table \ref{table:multiplicity} summarises the seven possible 1SBS variances of $\hat{\theta}$ given by \eqref{eq:that}, referred to as cases 1 to 7, which can result from the sample-dependent Strategy-III, in comparison with the traditional Strategy-I that requires an extra wave of SBS. Note that the two $\beta_{\Omega|s}$ discussed above actually lead to the same sampling variance, which are given as the two variations of case 5 in Table \ref{table:multiplicity}. The sample-dependent ancestor set $\beta_{\kappa|s}$ can only vary for a measurement unit with $|\beta_{\kappa}| >1$, which are $\kappa_2, \kappa_3, \kappa_4$ here; the sizes of $\beta_{\kappa|s}$ are given in italics for these three measurement units in Table \ref{table:multiplicity}. Finally, the coefficient of variation (CV) of the observed number of measurement units $|\Omega_s|$ is also given, which is seen to be correlated with $V(\hat{\theta})$ here. 

\begin{table}[ht]
\begin{center}
\caption{Some details of $1$SBS by Strategy-III in cases 2 and 5 (Table \ref{table:multiplicity}).}
\begin{tabular}{l|lr|lr} \toprule
 & \multicolumn{2}{c}{Case 2}  & \multicolumn{2}{|c}{Case 5} \\ %\cline{2-5}
$s_0$ &  $\Omega_s$ & $\hat{\theta}$  &  $\Omega_s$ & $\hat{\theta}$ \\ \hline 
$\{ i_1, i_2\}$ & $\{\kappa_1,\kappa_2,\kappa_3,\kappa_5\}$ & 6.4 & $\{\kappa_1,\kappa_2,\kappa_3,\kappa_5\}$ &	7.2  \\
$\{ i_1, i_3\}$ & $\{\kappa_1,\kappa_2,\kappa_3,\kappa_4\}$ & 6.4 &	 $\{\kappa_1,\kappa_2,\kappa_3\}$ & 5.2	\\
$\{ i_1, i_4\}$ & $\{\kappa_1,\kappa_2,\kappa_3,\kappa_6,\kappa_7\}$ & 8.4 & 	$\{\kappa_1,\kappa_2,\kappa_3,\kappa_4,\kappa_6,\kappa_7\}$ & 11.2 \\
$\{ i_2, i_3\}$ & $\{\kappa_2,\kappa_3,\kappa_4,\kappa_5\}$ & 6.4 &  $\{\kappa_5\}$ & 2.0 \\
$\{ i_2, i_4\}$ & $\{\kappa_2,\kappa_3,\kappa_5,\kappa_6,\kappa_7\}$ & 8.4 & $\{\kappa_3,\kappa_4,\kappa_5,\kappa_6,\kappa_7\}$ & 9.2 \\
$\{ i_3, i_4\}$ & $\{\kappa_4,\kappa_6,\kappa_7\}$ & 6.0 & $\{\kappa_3,\kappa_4,\kappa_6,\kappa_7\}$ & 7.2 \\ \hline 
$E(\hat{\theta})$ &  &  7.0 & & 7.0 \\ 
$V(\hat{\theta})$ &  & 1.000 & & 8.467 \\ \bottomrule
\end{tabular} \label{table:conditionalG} \end{center} \end{table}

The sampling variance of Strategy-III is lowest in case 2, where it is even more efficient that Strategy-I that requires an extra wave of SBS. The collection $\beta_{\Omega|s}$ in this case differs to $\{ \beta_{\kappa} : \kappa\in \Omega\}$ only in terms of $\kappa_3$ and $\kappa_4$, where $i_4$ is excluded from $\beta_{\kappa_3}$ and $\beta_{\kappa_4}$. The sampling variance of Strategy-III is highest in case 5, for which the two collections  $\beta_{\Omega|s}$ have been given in Figure \ref{fig:multiplicity}. Table \ref{table:conditionalG} compares the details for these two cases given all the six possible initial samples $s_0$, where $\beta_{\Omega|s}$ in the middle of Figure \ref{fig:multiplicity} is used for illustrating case 5 here. The association between $|\Omega_s|$ and $\hat{\theta}$ and its effect on the sampling variance are quite obvious in this SBS problem. Another indicator for the sampling variance of Strategy-III is the variability in the number of measurement units associated with each sampling unit given the collection $\beta_{\Omega|s}$, which is
\[
\nu_i = |\{ \kappa\in \Omega : i\in \beta_{\kappa|s} \}|
\]
We have $(\nu_{i_1}, \nu_{i_2}, \nu_{i_3}, \nu_{i_4}) = (3,3,1,2)$ in case 2, which has much less variability compared to $(\nu_{i_1}, \nu_{i_2}, \nu_{i_3}, \nu_{i_4}) = (3,1,0,4)$ or $(1,3,0,4)$ in case 5.

\subsection{Adaptive Cluster Sampling}

\emph{Adaptive cluster sampling (ACS)} from finite populations starts with an initial sample and \emph{adaptively} adds units to the current sample if they satisfy a criterion specified in advance. \citet{Thompson1990,Thompson1991} considers sampling from $U$ that consists of a set of spatial grids over a given area, where each grid is associated with an amount of a species of interest, denoted by $y_i$ for $i\in U$. The initial grid sample $s_0$ may be draw directly from $U$ or via some primary sampling units as in two-stage ACS. Given any $i\in s_0$, one would survey all its neighbour grids (in four directions if possible) \emph{only if} $y_i$ exceeds a threshold value. The observation is repeated for all the neighbour grids, which may or may not generate further grids to be surveyed. The process is terminated, when the last observed grids are all below the threshold. The interest is to estimate the total $\theta = \sum_{i\in U} y_i$. 

By recasting finite-population ACS as SBS from graphs, we explain below how the proposed strategies for SBS can extend the approach to ACS proposed by \citet{Thompson1990}. For illustration we use the ``small'' example of \citet{Thompson1990}, where the population $U$ consists of 5 spatial grids with $y$-values $\{ 1, 0, 2, 10, 1000\}$ and contiguity as represented by the edges in the top graph in Figure \ref{fig:ACS}. Notice that, following \citet{Thompson1990}, we simply denote each node (or grid) by its $y$-value. 

\begin{figure}[ht]
\centering
\begin{tikzcd}
1 \arrow[r, dash] & 0 \arrow[r, dash] & 2 \arrow[r, dash] & 10 \arrow[r, dash] & 1000 \\
1  & 0 & 2 & 10 \arrow[l] \arrow[r] & 1000 \arrow[l]
\end{tikzcd} 
\caption{Graph for ACS (top) or forward incident SBS (bottom).} \label{fig:ACS}
\end{figure}

ACS from the top graph in Figure \ref{fig:ACS} is the same as $\infty$SBS from the bottom graph by the forward (rather than reciprocal) incident observation procedure. A node with $y$-value below the threshold may be called a \emph{terminal node} because it does not lead to any other nodes. A terminal node is called an \emph{edge node}, if it is adjacent to at least one \emph{non-terminal network (NTN)}, which consists of a cluster of connected non-terminal nodes whose $y$-values all exceed the threshold, such as $\{ 10, 1000\}$ in Figure \ref{fig:ACS}. All the nodes in an NTN are observed under $\infty$SBS iff at least one of them is selected in $s_0$; observing an NTN would lead one to observe all its edge nodes, but not the other way around. 

Thus, all the nodes belonging to the same NTN have this NTN as their SBS ancestors, such as $\beta_{10} = \beta_{1000} = \{ 10, 1000\}$ in Figure \ref{fig:ACS}. For any terminal node $i$ that is not an edge node, its ancestor set is simply itself, $\beta_i = \{ i\}$, such as $\beta_1 = \{ 1\}$ and $\beta_0 = \{0\}$. The ancestor set of an edge node $i$ includes itself and all its adjacent NTNs, such as $\beta_2 = \{ 2, 10, 1000\}$. Notice that for $\theta$ as the total of $y_i$ over $U$, the motif of interest is node and no distinction between $\Omega$ and $U$ is necessary here. 

For Strategy-I (Section \ref{sec:I}), notice that one cannot observe $\beta_i$ of any edge node $i$ by additional waves of observation, unless all its adjacent NTNs intersect the initial $s_0$, because additional waves cannot progress from $i$. \citet{Thompson1990} proposes to modify the estimator \eqref{eq:that}, whereby an edge node is used in estimation \emph{only if} it is selected in $s_0$ directly, the probability of which can be calculated, but not when it is observed via its adjacent NTN. Denote this as Strategy-I$^*$ below due to the modified estimator $\hat{\theta}^*$. 

There is a strong connection to Strategy-II (Section \ref{sec:II}) using the estimator \eqref{eq:that}, where one can let $\beta_i^* = \{ i\}$ be the restricted ancestor set of any terminal node $i$, whether or not it is an edge node, which is possible since $U$ (or $\Omega$) is known. One can let $\beta_i^* = \beta_i$ for any non-terminal node, which can be observed under $\infty$SBS whenever $i$ is sampled.   

As a matter of fact one obtains the same estimate under the Strategy-II or I$^*$. The difference is that $\hat{\theta}$ would be unchanged by Rao-Blackwellisation, whereas $\hat{\theta}^*$ would; as will be illustrated below.

Finally, the sample-dependent Strategy-III (Section \ref{sec:III}) can be applied with $\beta_{i|s} =\beta_i$ for any non-edge node $i$. For an edge node, i.e. 2 in this example, the ancestors can e.g. be chosen as follows:
\begin{itemize}
\item[(a)] if $s_0\cap \{2,10,1000\}=\{2\}$, use $\beta_{2|s}=\beta_2^*=\{2\}$;
\item[(b)] if $s_0\cap \{10,1000\}\neq \emptyset$, use $\beta_{2|s} =\beta_2 =\{2,10,1000\}$. 
\end{itemize}

\begin{table}[ht]
\centering
\caption{SBS strategies for ACS from 1 --- 0 --- 2 --- 10 --- 1000.}
\begin{tabular}{l | lr | lr | lr} \toprule
& \multicolumn{2}{c}{I$^*$} & \multicolumn{2}{|c|}{II} & \multicolumn{2}{c}{III, $\beta_{2 | s}=\{2,10,1000\}$}\\ 
$s_0$ & Sample & $\hat{\theta}^*/5$ & Sample & $\hat{\theta}/5$ & Sample & $\hat{\theta}/5$ \\ \midrule
0,1 &	0,1 &	0.500 &	0,1	& 0.500 & 0,1	& 0.500 \\
0,2	& 0,2 &	1.000&	0,2&	1.000 & 0,2&	0.444 \\
0,10 & 	0,10,\emph{2},1000&	288.571&	0,10,1000&	288.571 & 0,10,2,1000&	289.016 \\
0,1000&	0,1000,\emph{2},10&	288.571&	0,1000,10&	288.571 & 0,1000,2,10&	289.016\\
1,2&	1,2	&1.500&	1,2&	1.500 & 1,2&	0.944 \\
1,10&	1,10,\emph{2},1000&	289.071&	1,10,1000&	289.071 & 	1,10,2,1000&	289.516\\
1,1000&	1,1000,\emph{2},10&	289.071&	1,1000,10&	289.071 & 1,1000,2,10&	289.516\\
2,10&	2,10,1000&	289.571&	2,10,1000&	289.571 & 2,10,1000&	289.016\\
2,1000&	2,1000,10&	289.571&	2,1000,10&	289.571 & 2,1000,10&	289.016\\
10,1000	&10,1000,\emph{2}&	288.571	&10,1000&	288.571 & 10,1000,2&	289.016\\ \hline
\multicolumn{1}{l}{Variance} & \multicolumn{2}{|r|}{17418.41} & \multicolumn{2}{r}{17418.41} & \multicolumn{2}{|r}{17482.35}\\ \bottomrule
\end{tabular} \label{table:ACS} \end{table}

For the numerical details of the various strategies in Table \ref{table:ACS}, we let the initial sample size be $n=2$ and select $s_0$ from $U$ by SRS, where the adaptive threshold is $y_i > 5$. For the Strategy-I$^*$, the edge node $2$ is given in italic in the five samples where it is observed but unused for estimation. The probability for this is $2/5$, which is just the sampling probability of the node 2 under Strategy-II. Otherwise, the observed sample of nodes are always the same under both the strategies, as well as the estimate. Nevertheless, the two differ regarding the Rao-Blackwell method. The difference hinges on the last three cases here, where the sample $\{ 2, 10, 1000\}$ is the same but $2$ is unused by Strategy-I$^*$ when $s_0 = \{ 10, 1000\}$. The Rao-Blackwell estimate in these three cases is $289.238$. In contrast, the estimate by Strategy-II is unchanged by Rao-Blackwellisation, because the sample used for estimation differs from $s_0 = \{ 10, 1000\}$ to that from $s_0 = \{ 2, 10\}$ or $\{ 2, 1000\}$.

For the sample-dependent Strategy-III, there exist two different ancestor sets for the edge node 2, depending on how it is actually observed. The results with $\beta_{2| s}=\{2\}$, given either $s_0 =\{0,2\}$ or $\{1,2\}$, are the same as those from Strategy-II, which are therefore omitted from Table \ref{table:ACS}. The results with $\beta_{2| s}=\{2, 10, 1000\}$ given $s_0\cap \{ 10,1000\}\neq \emptyset$ are presented. The unconditional variance of $\hat{\theta}$ by Strategy-III is $17468.14$, averaged over the two distinct collections $\beta_{\Omega|s}$ depending on $\beta_{2|s}$. It is less efficient than Strategy-II here, because the sampling probability of the edge node 2 is lower or equal under Strategy-II, which is more reasonable given that an edge node by definition has a relatively small $y$-value below the threshold. This can be easily amended under Strategy-III by setting $\beta_2^*$ to be a proper subset of $\beta_{2|s} =\{2, 10, 1000\}$.

\section{SBS for lower-order subgraphs} \label{sec:motif} 

Instead of node totals as in finite-population sampling, we now consider the totals of the other motifs defined for lower-order subgraphs in Figure \ref{fig:motifs}, i.e. edge ($\mathcal{K}_2$), 2-star ($\mathcal{S}_2$), triangle ($\mathcal{K}_3$), 4-cycle ($\mathcal{C}_4$), 4-clique ($\mathcal{K}_4$), 3-star ($\mathcal{S}_3$) and 3-path ($\mathcal{P}_3$). First, in Section \ref{sec:random}, the Strategies-II and III are illustrated for a simulated population graph and compared to graph sampling by induced observation. Next, in Section  \ref{sec:extreme}, some other graphs are used to further explore how the relative graph sampling efficiency might vary.

\subsection{A population graph with 50 nodes and 79 edges} \label{sec:random}  

\begin{figure}[ht]
\centering
\includegraphics[scale=0.75]{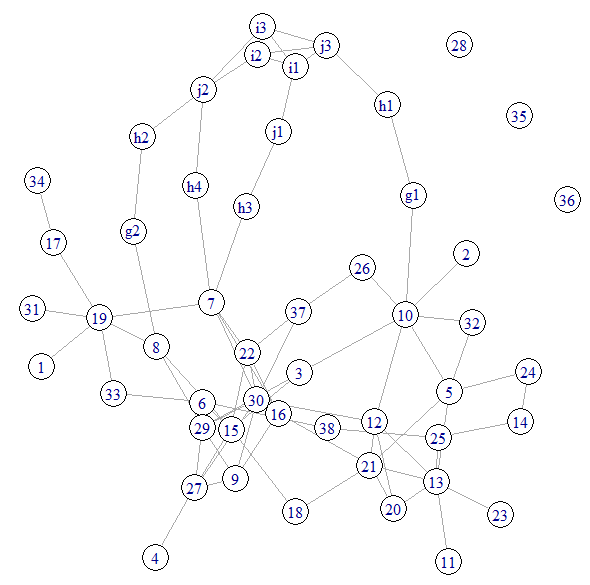}
\caption{A population graph with 50 nodes and 79 edges.}
\label{fig:popgraph}
\end{figure}

Figure \ref{fig:popgraph} shows a population graph with $50$ nodes and $79$ edges, the top central part of which has been given as Figure \ref{fig:TSBS} earlier. The graph totals of interest here are
\[
(\theta_{\mathcal{K}_2}, \theta_{\mathcal{S}_2}, \theta_{\mathcal{K}_3}, \theta_{\mathcal{C}_4}, \theta_{\mathcal{K}_4}, \theta_{\mathcal{S}_3}, \theta_{\mathcal{P}_3}) = (79, 207,  21, 10, 4, 161, 458).
\] 
 
\subsubsection{Some computational details} \label{sec:detail}  

Before presenting the results of relative efficiency in Section \ref{sec:results}, let us first explain some computational details of the SBS strategies by an example in terms of $\mathcal{K}_3$, $\mathcal{C}_4$ and $\mathcal{K}_4$.

\begin{figure}[ht]
\centering
\includegraphics[width=15cm, height=15cm]{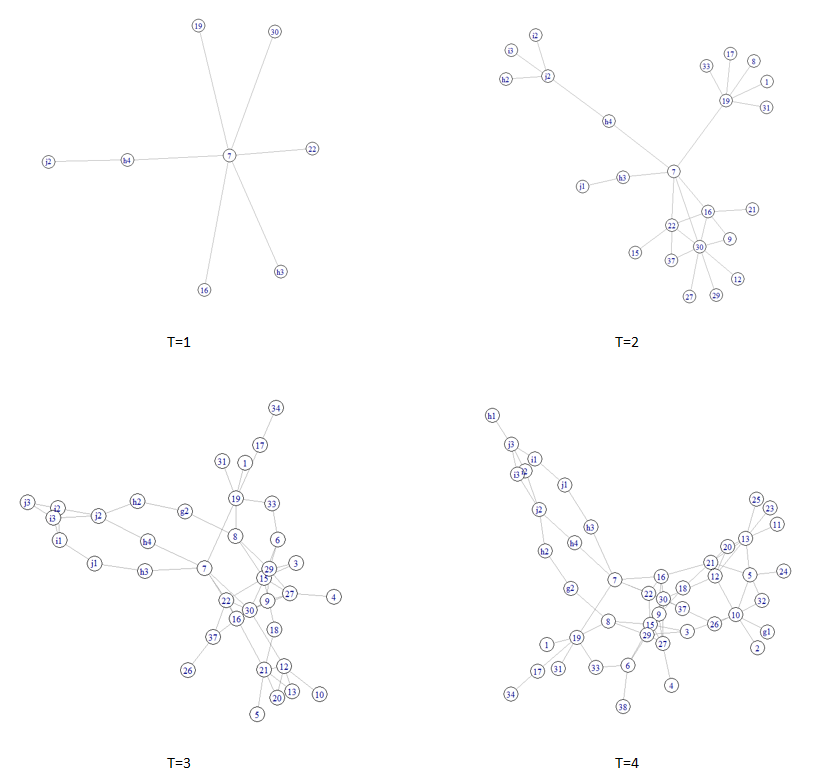}
\caption{Sample graphs by $T$SBS for $T=1,...,4$ given initial sample $s_0 =\{h_4,7\}$.}
\label{fig:GsTSBS}
\end{figure}

Let $s_0=\{h_4,7\}$ be the initial sample by SRS from $U$. The resulting sample graphs by $T$SBS for $T=1, 2, 3, 4$ are given in Figure \ref{fig:GsTSBS}. None of the three motifs $\mathcal{K}_3$, $\mathcal{C}_4$ and $\mathcal{K}_4$ is observed by $T=1$. Only the nodes $\{28,35,36,14\}$ are yet unobserved by the 4th wave, where the first three have degree zero (i.e. isolated nodes). All the edges incident to the nodes $s_0 \cup \cdots \cup s_3$ are observed, but not all of those incident to $s_4$.  

For any observed sample motif $\kappa$ by given $T$, let $|\beta_{\kappa}|$ be the multiplicity of its $T$SBS ancestor set, and let $|\beta_\kappa^*|$ or $|\beta_{\kappa|s}|$ be that by Strategy-II or III. The motif $\kappa$ can be used for estimation by Strategy-II if $s_0 \cap \beta_{\kappa}^* \neq \emptyset$, whereas it can be used by Strategy-III if $s_0 \cap \beta_{\kappa|s} \neq \emptyset$. Given SRS of the initial $s_0$, the probability \eqref{eq:pik} is given by
\[
\pi_{(\kappa)} = 1- \mathcal{C}(N-m_{\kappa}, n)/\mathcal{C}(N, n) 
\]
where $N =50$, $n = 2$ and $m_\kappa = |\beta_\kappa^*|$ or $|\beta_{\kappa|s}|$ depending on the adopted strategy.

\begin{table}[ht]
\centering
\caption{Multiplicity of triangles $\mathcal{K}_3$, 4-cycles $\mathcal{C}_4$ and 4-cliques $\mathcal{K}_4$ observed under $T$SBS with initial $s_0 = \{ h_4,7\}$. All $T$SBS ancestors $|\beta_{\kappa}|$, restricted $|\beta_{\kappa}^*|$ for Strategy-II (``--'' if $s_0\cap \beta_{\kappa}^* = \emptyset$), sample-dependent $|\beta_{\kappa|s}|$ for Strategy-III (``--'' if $s_0\cap \beta_{\kappa|s} = \emptyset$).}
\begin{tabular}{llrrrr} \toprule
Motif $\kappa$ & $T$SBS & Nodes $M_\kappa$ & $|\beta_\kappa^*|$ & $|\beta_{\kappa|s}|$ & $|\beta_{\kappa}|$ \\ \midrule
\multirow{6}{*}{$\mathcal{K}_3$} & \multirow{6}{*}{$T=2$} & $\{7,16,22\}$ & 4 & 4 & 4 \\ 
& & $\{7,16,30\}$ & 4 & 5 & 5 \\ 
& & $\{7,22,30\}$ & 4 & 5 & 5 \\
& & $\{16,30,22\}$ & 4 & 6 & 6 \\ 
& & $\{16,30,9\}$ & -- & 5 & 7 \\
& & $\{22,30,37\}$ & -- & 5 & 5 \\ \bottomrule 
\multirow{8}{*}{$\mathcal{C}_4$} & \multirow{8}{*}{$T=3$} & $\{15,27,29,3\}$  & -- & 14 & 16 \\ 
& & $\{15,29,3,8\}$  & -- & 14 & 15 \\
& & $\{15,29,6,8\}$  & -- & 14 & 15 \\
& & $\{15,27,29,6\}$  & -- & 16 & 17 \\
& & $\{12,16,21,30\}$  & -- & 16 & 16 \\
& & $\{15,27,29,8\}$  & -- & 17 & 17 \\
& & $\{15,22,27,30\}$  & -- & 17 & 17 \\ \bottomrule
\multirow{7}{*}{$\mathcal{K}_4$} & \multirow{1}{*}{$T=2$} & $\{7,16,22,30\}$ & 4 & 4 & 4 \\ \cline{2-6}
& \multirow{2}{*}{$T=3$} & $\{7,16,22,30\}$ & 14 & 15 & 15 \\ 
& & $\{27,29,30,9\}$ & 14 & 14 & 14 \\ \cline{2-6}
& \multirow{4}{*}{$T=4$} & $\{7,16,22,30\}$ & 14 & 31 & 31 \\ 
& & $\{27,29,30,9\}$ & 14 & 27 & 27 \\ 
& & $\{12,13,20,21\}$ & -- & 25 & 28 \\ 
& & $\{i_1,i_2,i_3,j_3\}$ & -- & 11 & 12 \\ \bottomrule
\end{tabular} \label{table:detail}
\end{table}

Take first the triangle $\mathcal{K}_3$, the number of which is $\theta = 21$ in Figure \ref{fig:popgraph}. Table \ref{table:detail} lists the six triangles that are observed by 2SBS from $s_0=\{h_4,7\}$ and their multiplicity $|\beta_{\kappa}|$. The estimate of $\theta$ is 25.8 by Strategy-II and 31.7 by Strategy-III.
\begin{itemize}[leftmargin=6mm]
\item[(a)] For Strategy-II, as explained in Section \ref{sec:II}, Lemma \ref{lemma:combineddefault} yields $\beta_{\kappa}^* = M_{\kappa}$ for any triangle $\kappa$ under 2SBS, whereas Lemma \ref{lemma:simpleextension} yields additionally any node $i$ that is adjacent to all the nodes $M_{\kappa}$, such as the node 30 for $M_{\kappa} = \{ 7, 16, 22\}$ in Table \ref{table:detail}. Hence, we have $\beta_{\kappa}^* = \{ 7, 16, 22, 30\}$ and $s_0 \cap \beta_{\kappa}^* = 7$ for all the first four triangles. No such additional nodes exist for the fifth or sixth triangle, for which we have $\beta_{\kappa}^* = M_{\kappa}$ and $s_0 \cap \beta_{\kappa}^* = \emptyset$, such that neither is used in estimation although both are observed.

\item[(b)] For the sample-dependent Strategy-III, we only need to identify the 2SBS ancestors in the sample graph $T=2$ in Figure \ref{fig:GsTSBS} directly, such as $\beta_{\kappa|s} = \{ 7, 16, 22, 30\}$ for $M_{\kappa} = \{ 7, 16, 22\}$, $\beta_{\kappa|s} = \{ 7, 16, 30, 22, 9\}$ for  $M_{\kappa} = \{ 7, 16, 30\}$, and so on.
\end{itemize}

Consider next the 4-cycle $\mathcal{C}_4$, for which $\theta = 10$ in Figure \ref{fig:popgraph}. Given $s_0=\{h_4,7\}$, no 4-cycle is observed by $T=2$ while seven of them are observed by $T=3$ (Table \ref{table:detail}). The estimate $\hat{\theta}$ does not exist by Strategy-II, while it is 13.4 by Strategy-III.
\begin{itemize}[leftmargin=6mm]
\item[(c)] Under 3SBS, only Lemma \ref{lemma:combineddefault} is applicable for Strategy-II, which yields $\beta_{\kappa}^*$ as the union of $M_{\kappa}$ and any other node $i$ with observation distance $d_{i,\kappa} \leq t_{\kappa} = 2$. However, no such additional nodes (outside $M_{\kappa})$ exist for these seven 4-cycles, such that $\beta_{\kappa}^* = M_{\kappa}$. Since $s_0 \cap \beta_{\kappa}^* = \emptyset$ for all the 4-cycles here, none of them can be used by Strategy-II.

\item[(d)] For the sample-dependent Strategy-III, we identify the 3SBS ancestors in the sample graph by $T=3$ in Figure \ref{fig:GsTSBS} directly. This yields $|\beta_{\kappa}^*|$ either equal to $|\beta_{\kappa}|$ or almost so for the first seven 4-cycles; the probability \eqref{eq:pik} is $0.4857$ for the first three of them, $0.5420$ for the next two and $0.5690$ for the last two. 
\end{itemize}

Finally, there are only four 4-cliques $\mathcal{K}_4$ in Figure \ref{fig:popgraph}. Given $s_0=\{h_4,7\}$, all of them are observed by $T=4$ (Table \ref{table:detail}). 
\begin{itemize}[leftmargin=6mm]
\item[(e)] Only the motif $\kappa$ with $M_{\kappa} = \{7, 16, 22, 30\}$ is observed by $T=2$, for which we have $\beta_{\kappa}^*=\beta_{\kappa|s}=M_{\kappa}$, such that $\pi_{(\kappa)} = 0.1551$ and $\hat{\theta} = 6.4$ either by Strategy-II or III.  

\item[(f)] One more 4-clique is observed by $T=3$. For Strategy-II by Lemma \ref{lemma:combineddefault}, $\beta_{\kappa}^*$ is the union of $M_{\kappa}$ and any node that is adjacent to at least one of $M_{\kappa}$. By Strategy-III, one additional node 8 is included in $\beta_{\kappa|s}$ for $M_{\kappa} = \{7, 16, 22, 30\}$, which is two steps away from any node in $M_{\kappa}$. The probability $\pi_{(\kappa)}$ is $0.4857$ if $m_{\kappa} = 14$ or $0.5143$ if $m_{\kappa} =15$, yielding $\hat{\theta}=4.1$ or 4.0 either by Strategy-II or III. 

\item[(g)] The last two 4-cliques observed by $T=4$ are not used by Strategy-II, since $s_0 \cap \beta_{\kappa}^* =\emptyset$ for them. Hence, $\hat{\theta} = 4.1$ as by the same strategy under 3SBS. In contrast, all the observed motifs are used by Strategy-III, yielding $\hat{\theta} = 6.3$. 
%The relatively large difference to the estimate 4.0 observed by the same strategy under 3SBS seems to suggest that the probabilities $\pi_{(\kappa)}$ mattered more than the number of observed sample motifs here.
\end{itemize}

\subsubsection{Results} \label{sec:results}

Table \ref{table:relvarG1} provide some results of relative efficiency for totals of all the motifs in Figure \ref{fig:popgraph} mentioned at the beginning of Section \ref{sec:motif}. $T$SBS either by Strategy-II or III is compared to graph sampling by induced observation following initial SRS; the sample size for the latter is set to the expected node sample size by the corresponding $T$SBS.

\begin{table}[ht]
\centering
\caption{Relative efficiency of $T$SBS from Figure \ref{fig:popgraph} by Strategy-II or III following initial SRS of $s_0$ with $|s_0| =2$, against graph sampling by induced observation following initial SRS with sample size $n = 7, 19, 32, 40$ for $T = 1, 2, 3, 4$.}
\begin{tabular}{llrrrrrrr} \toprule
& Strategy & $\mathcal{K}_2$ & $\mathcal{S}_2$ & $\mathcal{K}_3$ & $\mathcal{K}_4$ & $\mathcal{C}_4$ & $\mathcal{S}_3$ & $\mathcal{P}_3$ \\ \midrule
\multirow{3}{*}{$T=1$} & \multirow{1}{*}{III}    &0.26&--&--&--&--&--&--\\					
& \multirow{1}{*}{II, Corollary \ref{cor:TforbetaK}}  &0.26&--&--&--&--&--&--\\	
&\multirow{1}{*}{II, Lemma \ref{lemma:combineddefault}}  &0.26&--&--&--&--&--&--\\ \bottomrule
\multirow{4}{*}{$T=2$} & \multirow{1}{*}{III}  &1.63&	0.76&	0.50&	0.08&	0.16&	1.31& 0.71\\ 
& \multirow{1}{*}{II, Corollary \ref{cor:TforbetaK}}  &2.18&--&--&--&--&--&--\\	
&\multirow{1}{*}{II, Lemma \ref{lemma:combineddefault}}  &2.18&0.74&0.57&0.08&0.16&1.39&0.69\\
&\multirow{1}{*}{II, Lemma \ref{lemma:simpleextension}}  &2.25&0.76&0.57&0.08&0.16&1.39&0.70\\ \bottomrule
\multirow{3}{*}{$T=3$} & \multirow{1}{*}{III} & 5.13 &	1.78 &	0.61 &	0.24 &	0.29 &	1.25 & 1.12\\ 
& \multirow{1}{*}{II, Corollary \ref{cor:TforbetaK}}  &5.80&3.38&2.54&0.94&1.25&8.03&3.44\\	
&\multirow{1}{*}{II, Lemma \ref{lemma:combineddefault}}  &5.80&3.38&0.93&0.26&1.25&1.42&1.73\\ \bottomrule
\multirow{4}{*}{$T=4$} & \multirow{1}{*}{III} &5.82&	2.21&	0.85&	0.41&	0.24&	1.51 & 1.22\\ 
& \multirow{1}{*}{II, Corollary \ref{cor:TforbetaK}}  &14.42&8.79&6.98&3.45&3.91&23.23&9.14\\	
&\multirow{1}{*}{II, Lemma \ref{lemma:combineddefault}}  &14.42&4.91&2.54&0.94&1.10&2.24&3.51\\	
&\multirow{1}{*}{II, Lemma \ref{lemma:simpleextension}}  &13.19&4.96&2.27&0.94&1.08&2.25&3.75 \\ \bottomrule
\end{tabular} \label{table:relvarG1}
\end{table}

For these results we let $|s_0|=2$ by SRS for $T$SBS with $T\leq 4$, where a large part of the graph may already have been observed by $T=4$ as illustrated above in Section \ref{sec:detail}. Given the observation diameters $(1,2,2,2,2,3,3)$ of $(\mathcal{K}_2, \mathcal{S}_2, \mathcal{K}_3, \mathcal{K}_4, \mathcal{C}_4, \mathcal{S}_3, \mathcal{P}_3)$, we need $T\geq 2$ to apply the SBS strategies in Section \ref{sec:theory} except for edge $\mathcal{K}_2$. The restricted ancestor set $\beta_{\kappa}^*$ for Strategy-II is given by Corollary \ref{cor:TforbetaK}, Lemma \ref{lemma:combineddefault} or Lemma \ref{lemma:simpleextension} (in case $T$ is even), where Corollary \ref{cor:TforbetaK} is only applicable for edge under 2SBS. An observed sample motif $\kappa$ is used by Strategy-II if $s_0 \cap \beta_{\kappa}^* \neq \emptyset$, or by Strategy-III if $s_0 \cap \beta_{\kappa|s} \neq \emptyset$. 

The unconditional variance of Strategy-III cannot be evaluated here due to the large number of possible $\beta_{\Omega|s}$. In  Table \ref{table:relvarG1} we report the conditional variance given the following $\beta_{\Omega|s}$. Let a hypothetical sequence of $s_0$ be $\{ 1, 2\}$, $\{ 1, 3\}$, $\dots$, $\{1, 50\}$, $\{ 2, 3\}$, $\dots$, $\{ 2, 50\}$, $\dots$, $\{ 49, 50\}$, where the nodes 39 to 50 correspond to  $i_1$, $i_2$, $i_3$, $j_1$, $j_2$, $j_3$, $h_1$, $h_2$, $h_3$, $h_4$, $g_1$ and $g_2$. Going through $s_0$ one by one in this sequence, we let $\beta_{\kappa |s}$ be given according to the first sample graph where $\beta_{\kappa |s} \cap s_0 \neq \emptyset$ for each relevant motif $\kappa$, until we obtain $\beta_{\Omega|s}$. Although this conditional variance differs to the unconditional variance of Strategy-III, the comparisons to be discussed below are robust against this difference. 

Given the observed sample of motifs $\Omega_s$ under graph sampling by induced observation following initial SRS of size $n$ (from $N$ nodes), the Horvitz-Thompson estimator \citep{Horvitz1952} of $\theta$ defined by \eqref{theta} is given by 
\[
\hat{\theta}_{HT}=\sum_{\kappa \in \Omega_s} \frac{y_\kappa}{\pi_{(\kappa)}} 
\]
where
\[
\pi_{(\kappa)} = \mathcal{C}(N-|M_\kappa|, n-|M_\kappa|)/\mathcal{C}(N, n)
\]
for any $\kappa\in \Omega$, and its sampling variance is
\[
V(\hat{\theta}_{HT}) = \sum_{\kappa\in \Omega} \sum_{l\in \Omega} \big( \frac{\pi_{(\kappa\ell)}}{\pi_{(\kappa)} \pi_{(\ell)}} - 1 \big) y_{\kappa} y_{\ell}
\]
where
\[
\pi_{(\kappa\ell)} = \mathcal{C}(N-|M_\kappa\cup M_\ell|, n-|M_\kappa\cup M_\ell|)/\mathcal{C}(N, n)
\]
for $\kappa\neq \ell \in\Omega$ and $\pi_{(\kappa\kappa)} = \pi_{(\kappa)}$ for any $\kappa\in \Omega$.

It can be seen from Table \ref{table:relvarG1} that, for Strategy-II, Lemma \ref{lemma:combineddefault} can yield great efficiency gains over Corollary \ref{cor:TforbetaK}, but the difference to Lemma \ref{lemma:simpleextension} is little here. For $T\geq 3$, the sample-dependent Strategy-III can be much more efficient than Strategy-II, whereas graph sampling by the induced observation procedure is more efficient than SBS by Strategy-III for the star-like motifs $(\mathcal{S}_2, \mathcal{S}_3, \mathcal{P}_3)$ but less efficient for the clique-like motifs $(\mathcal{K}_3, \mathcal{K}_4, \mathcal{C}_4)$. 

The results for the edge total seems particularly instructive, where SBS is much more efficient than induced graph sampling with $T=1$ but much less efficient as $T$ increases. As $T$ increases, so does the variability in the multiplicity $m_{\kappa}$ (or sampling probability $\pi_{(\kappa)}$) over $\kappa\in \Omega$, which is not desirable here since $y_{\kappa} \equiv 1$ over $\Omega$. For example, by Strategy-III, the range of $\pi_{(\kappa)}$ for edges is $[0.0792,0.0792]$ with $T=1$, $[0.1176,0.4261]$ with $T=2$, $[0.1551,0.7551]$ with $T=3$ and $[0.3633,0.9551]$ with $T=4$. In contrast, the sample inclusion probability is the same for all the edges under induced graph sampling following initial SRS, i.e. 0.0171, 0.1396, 0.4049 or 0.6367 for $T=1, 2, 3$ or 4 here.

Of course, this also suggests that the relative efficiency of $T$SBS could improve if the value of interest $y_{\kappa}$ (associated with motif $\kappa$) tends to be higher in the denser parts of the graph (i.e. instead of $y_{\kappa} \equiv 1$ considered here). For example, the bottom part of the population graph in Figure \ref{fig:popgraph} is more dense than the other parts of the graph. If a node in this part is selected in $s_0$, we would observe relatively more edges (and other motifs) as $T$ increases, compared to induced graph sampling. If such variations of $\pi_{(\kappa)}$ are positively associated with $y_{\kappa}$, then the relative efficiency of $T$SBS would improve.

\subsection{Some further exploration} \label{sec:extreme}

\begin{figure}[ht]
\centering
\includegraphics[scale=0.75]{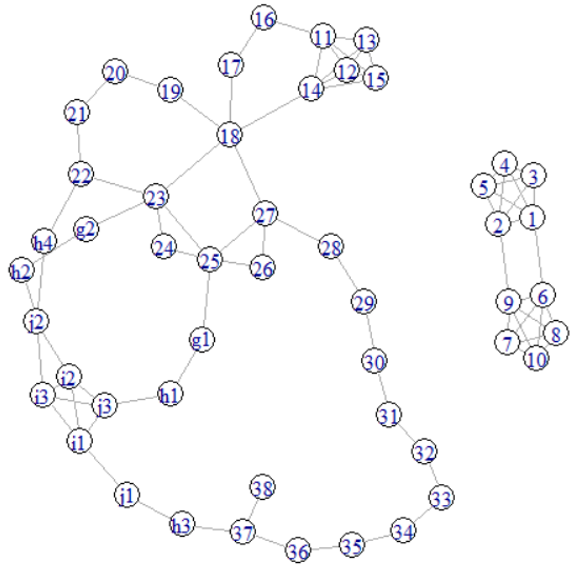}
\caption{Another population graph with 50 nodes and 79 edges.}
\label{fig:anothergraph}
\end{figure}

The results in Section \ref{sec:results} suggest that, given $y_{\kappa} \equiv 1$ for all $\kappa \in \Omega$, the relative efficiency of $T$SBS may be improved if the nodes in the population graph have a more even degree distribution. Figure \ref{fig:anothergraph} shows another population graph with the same number of nodes and edges as that in Figure \ref{fig:popgraph}, but apparently with a much more even degree distribution; it mainly consists of cycles and some small cliques (up to order 5). 

\begin{table}[!h]
\centering
\caption{Relative efficiency of $T$SBS from Figure \ref{fig:anothergraph} by Strategy-II or III following initial SRS of $s_0$ with $|s_0| =2$, against graph sampling by induced observation following initial SRS with sample size $n=8,14,21, 27$ for $T=1,2,3,4$.}
\begin{tabular}{llrrrrrrr} \toprule
$T$SBS & Strategy & $\mathcal{K}_2$ & $\mathcal{S}_2$ & $\mathcal{K}_3$ & $\mathcal{K}_4$ & $\mathcal{C}_4$ & $\mathcal{S}_3$ & $\mathcal{P}_3$ \\ \midrule
\multirow{3}{*}{$T=1$} & \multirow{1}{*}{III}   & 0.17&--&--&--&--&--&--\\ 					
&\multirow{1}{*}{II, Corollary \ref{cor:TforbetaK}} &0.17&--&--&--&--&--&-- \\ 
&\multirow{1}{*}{II, Lemma \ref{lemma:combineddefault}} &0.17&--&--&--&--&--&-- \\ \bottomrule
\multirow{4}{*}{$T=2$} & \multirow{1}{*}{III}  &0.72&	0.19&	0.24&	0.04&	0.02&	0.31&	0.23 \\ 
& \multirow{1}{*}{II, Corollary \ref{cor:TforbetaK}} &0.62&	--&	--&	--&	--&	--&	-- \\ 
& \multirow{1}{*}{II, Lemma \ref{lemma:combineddefault}} &0.62&	0.22&	0.24&	0.04&	0.02&	0.31&	0.23 \\
& \multirow{1}{*}{II, Lemma \ref{lemma:simpleextension}} &0.66&	0.22&	0.24&	0.04&	0.02&	0.31&	0.23 \\ \bottomrule
\multirow{3}{*}{$T=3$} & \multirow{1}{*}{III} &2.39&	0.35&	0.60&	0.17&	0.03&	0.28&	0.19\\ 
& \multirow{1}{*}{II, Corollary \ref{cor:TforbetaK}} &1.78&	0.53&	0.74&	0.18&	0.13&	1.27&	0.87\\ 
& \multirow{1}{*}{II, Lemma \ref{lemma:combineddefault}} &1.78&	0.53&	0.63&	0.17&	0.13&	0.34&	0.31\\ \bottomrule
\multirow{4}{*}{$T=4$} & \multirow{1}{*}{III} &4.98&	0.63&	1.05&	0.35&	0.08&	0.27&	0.29 \\ 
& \multirow{1}{*}{II, Corollary \ref{cor:TforbetaK}} &3.72&	1.15&	1.59&	0.51&	0.42&	3.00&	2.05 \\ 
& \multirow{1}{*}{II, Lemma \ref{lemma:combineddefault}} &3.72&	0.68&	1.36&	0.46&	0.08&	0.31&	0.63 \\ 
& \multirow{1}{*}{II, Lemma \ref{lemma:simpleextension}} &4.39&	0.70&	1.44&	0.47&	0.09&	0.31&	0.64 \\ \bottomrule
\end{tabular} \label{table:relvarG2}
\end{table}

Table \ref{table:relvarG2} presents the results of relative efficiency in parallel to Table \ref{table:relvarG1}, where the initial sample size for induced graph sampling is now $n=8,14,21$ or 27 for $T=1,2,3$ or 4. For Strategy-II, we observe the same pattern of relative efficiency among Corollary \ref{cor:TforbetaK}, Lemma \ref{lemma:combineddefault} and Lemma \ref{lemma:simpleextension}. Next, the differences between Strategy-III and Strategy-II (by Lemma \ref{lemma:combineddefault} or \ref{lemma:simpleextension}) have reduced considerably compared to those for Figure \ref{fig:popgraph}. Finally, except for the edge total, $T$SBS by Strategy-III is mostly more efficient than graph sampling by induced observation including the star-like motifs $(\mathcal{S}_2, \mathcal{S}_3, \mathcal{P}_3)$.

\begin{table}[ht]
\centering
\caption{Relative efficiency for edge total, $T$SBS following initial SRS of $s_0$ with $|s_0| =2$ against graph sampling by induced observation following initial SRS with sample size $n$.}
\begin{tabular}{cccc|cc} \toprule
\multicolumn{4}{c|}{50-cycle} & \multicolumn{2}{c}{(48-clique, 2-clique)} \\
$T=1$ & $T=2$ & $T=3$ & $T=4$ & $T=1$ & $T=2$ \\ 
$n=5$ & $n=9$ & $n=13$ & $n=16$ & $n=48$ & $n=48$  \\ \midrule
0.0012 & 0.0052 & 0.0478 & 0.0873 & 135.23 & 5.72 \\ \bottomrule
\end{tabular} \label{table:relvaradhoc}
\end{table}

\begin{table}[!h]
\centering
\caption{Edge total estimation in 50-cycle graph, variance $V_{II}$ by Strategy-II under $T$SBS  or $V_0$ by induced graph sampling following initial SRS of sample size $n$.}
\begin{tabular}{crrrrrrrrrr} \toprule
$T$ & 1 & 2 & 3 & 4 & 5 & 6 & 7 & 8 & 9 & 10 \\ \midrule
$V_{II}$  & 6 & 6& 21 & 21 & 36 & 36 & 50& 50 & 63 & 63 \\
$V_0$ & 5156 & 1186 & 445 & 243 & 142 & 85 & 52 & 32 & 19 & 13\\
$n$ & 5  & 9 & 13 & 16 &19 & 22 & 25 & 28 & 31 & 33\\ \bottomrule
\end{tabular} \label{tab:50cycle}
\end{table}

Focusing on the edge total, we now construct two more extreme population graphs: (i) a 50-cycle, and (ii) a two-component graph of a 48-clique and a 2-clique. Table \ref{table:relvaradhoc} presents the relative efficiency of $T$SBS by Strategy-II following SRS of $s_0$ with $|s_0|=2$, against induced graph sampling following initial SRS with comparable sample size $n$. The restricted ancestor set $\beta_{\kappa}^*$ is either given by Lemma \ref{lemma:combineddefault} if $T$ is odd or by Lemma \ref{lemma:simpleextension} if $T$ is even. Notice that for the second population graph (48-clique, 2-clique), we let $T =2$ at most, because the sample graph would not change after two waves no mater how $s_0$ is distributed between the two components of the graph.

Clearly, the relative efficiency of $T$SBS given $T\leq 4$ is considerably improved given a more even degree distribution in the population graph. Indeed, $T=1$ may be the most efficient choice of $T$ for edge total estimation in an $N$-cycle graph, because the number of observed sample edges would be a constant as long as the initial sample nodes are at least two steps away from each other. Whereas as $T$ increases, the chance would increase for the observed paths (from each initial sample  node) to overlap by varying extent, which would result in an increase of the sampling variance for edge total estimation. Table \ref{tab:50cycle} shows the sampling variance by Strategy-II under $T$SBS from the 50-cycle graph, for $1\leq T\leq 10$, as well as that of induced graph sampling following initial SRS of comparable sample sizes. It can be seen that the $T$SBS sampling variance is indeed the lowest at $T=1$. In comparison, the sampling variance of induced graph sampling decreases in a monotone fashion as $T$ increases. The relative efficiency is almost equal to one at $T=7$ in this case, afterwards induced graph sampling becomes more efficient than $T$SBS.

\section{Final remarks} \label{sec:final}

Graph sampling \citep{Zhang2021,ZhangPatone2017} provides a statistical approach to study real graphs, whether the graph is only partially observed in terms of the sample graph or if the graph is in principle known but various compressed views of it are desirable (compared to processing the whole graph). In the above we have developed a general theory for SBS from graphs. It encompasses the relevant works on probabilistic SBS from graphs or finite populations, thereby extending the scope of application of snowball sampling. In situations where the estimation objective is also achievable by other methods, such as induced graph sampling, the theory provides one with a greater choice of design-unbiased strategies and the means to improve graph sampling efficiency in practice. Below we shortly point out some topics for future research. 

Just like the estimator for ACS proposed by \citet{Thompson1990} can be viewed as a modification of the estimator \eqref{eq:that} under Strategy-I for SBS, other estimators are possible under the outlined SBS strategies. Indeed, \citet{PatoneZhang2022} develop a large class of design-unbiased incidence weighting estimators, which greatly extends the work of \citet{BirnbaumSirken1965} and includes the estimator \eqref{eq:that} as a special case. This broad class of estimators are applicable to SBS given any of the Strategies I-III to $\beta_{\kappa}^*$, although an investigation of such extensions of \eqref{eq:that} is beyond the scope of the current paper. Other unbiased or nearly unbiased estimators may be discovered in future.

As discussed at the end of Section \ref{sec:I}, $T$SBS ancestors can be generalised from nodes to sets of nodes. Provided one can handle the required graph  computations, this would introduce additional possibilities for improving the efficiency of graph sampling. 

The incident observation procedure of SBS considered in this paper is the same as that of breadth-first search in graphs, which includes \emph{all} the adjacent nodes that can be reached from a current node, possibly subjected to the direction of observation. In contrast, depth-first observation would include only one of these adjacent nodes, as in various random-walk related sampling methods \citep[e.g.][]{Thompson2006SM, Zhang2021, Zhang2022}. Hybrid procedures of breadth- and depth-first observation can be envisaged, such as adaptive web sampling \citep{Thompson2006}. Further development of hybrid graph sampling methods and associated strategies is another interesting topic.

Finally, although we have developed a general graph sampling theory for finite-order subgraphs, there are many \emph{other kinds} of graph parameters that are of great interest. For instance, various centrality and betweenness measures are important in network analysis \citep[e.g.][]{Newman2010}, which are defined locally for a given node but depend on the whole graph. Random-walk-like sampling methods for embedding words in a text to vectors \citep[e.g.][]{Mikolov2013} can be viewed as a related approach to similar graph parameters. Efficient sampling methods for such graph parameters constitute a large topic area.

\appendix 
\section{Appendix: Proofs} \label{proofs}

Lemma \ref{lemma:dik}.
\begin{proof} \label{proof-1}
If there is only one node, denoted by $j_0$, which is observed by the $\lambda_{i,\kappa}$-th wave from $i$, then all other nodes in $M_{\kappa}$ must have been observed before the $\lambda_{i,\kappa}$-th wave. Thus all edges among $M_{\kappa}$ are observed by the $\lambda_{i,\kappa}$-th wave at the latest, when $j_0$ is observed. If there are two or more nodes like $j_0$, requiring $\lambda_{i,\kappa}$ waves to be observed from $i$, one more wave is needed to observe any edges among them.   
\end{proof}

\noindent Lemma \ref{lemma:ancestralT}.
\begin{proof} \label{proof-2}
If $|M_{\kappa}| >1$, then applying $T-1$ waves of RIOP to $M_{\kappa}$ would identify all the nodes $\{ i\in U :\varphi_{i,\kappa}\leq T-1\}$. By Lemma \ref{lemma:dik}, any node $i$ with $\varphi_{i,\kappa} \geq T$ cannot be a $T$SBS ancestor of $\kappa$ now that $|M_{\kappa}| >1$. If $|M_{\kappa}|=1$, then applying at most $T$ waves of RIOP to $M_{\kappa}$ would identify all the nodes $\{ i\in U : d_{i,\kappa}\leq T \}$. 
\end{proof}

\noindent Corollary \ref{cor:TforbetaK}.
\begin{proof} \label{proof-c}
After $\kappa$ is observed from any $i \in \beta_{\kappa}^*$ by at most $t_{\kappa}$ waves, at most $t_{\kappa}-1$ or $t_{\kappa}$ waves are needed to observe all the other nodes in $\beta_{\kappa}^*$, depending on $|M_{\kappa}| >1$ or $|M_{\kappa}| =1$.  
\end{proof}

\noindent Lemma \ref{lemma:combineddefault}.
\begin{proof} \label{proof-3}
The motif $\kappa$ can be observed starting from any node in $\beta_{\kappa}^*$ after at most $T$ waves, such that $\beta_{\kappa}^*$ are all $T$SBS ancestors. It remains to show that any node $j\in \beta_{\kappa}^*$ can be identified as a $T$SBS ancestor after $T$ waves starting from any node in $\beta_{\kappa}^*$. 

In case $\check{\beta}_{\kappa} = \{ i\not \in M_{\kappa} : \varphi_{i,\kappa} \leq t_{geo} \}$ does not exist, $\tilde{\beta}_{\kappa} = \{ i\in U :d_{i,\kappa} \leq t_{\kappa} \}$ is a restricted set by Corollary \ref{cor:TforbetaK} and, starting from any node in $\tilde{\beta}_{\kappa}$, $G(M_{\kappa})$ is observed and connected, so that $\dot{\beta}_{\kappa} = \{ i\in M_{\kappa} : d_{i,\kappa} \leq T\}$ can be identified. Whereas starting from any node in $\dot{\beta}_{\kappa}$, $\tilde{\beta}_{\kappa}$ are all observed because $t_{\kappa} \leq T$, as well as $\dot{\beta}_{\kappa}$ by definition of $\dot{\beta}_{\kappa}$. 

In case $\check{\beta}_{\kappa}$ exists, then $\beta_{\kappa}^* = \dot{\beta}_{\kappa} \cup \check{\beta}_{\kappa}$, because $\tilde{\beta}_{\kappa} \subseteq \check{\beta}_{\kappa}$. This can be shown by proving that $\max_{i\in \check{\beta}_{\kappa}} d_{i,\kappa} \geq t_{\kappa}$, where $\max_{i\in \check{\beta}_{\kappa}} d_{i,\kappa} = \zeta_{\kappa}+\lfloor(T-\varphi_{\kappa})/2\rfloor$. This is obvious if $|M_{\kappa}| = 1$, in which case $\max_{i\in \check{\beta}_{\kappa}} d_{i,\kappa} = \lfloor T/2\rfloor = t_{\kappa}$. Otherwise, $t_{\kappa} = \lfloor (T+1)/2\rfloor$, whereas $\varphi_{\kappa}$ is either odd or even. If $\varphi_{\kappa}$ is odd, then $\varphi_{\kappa} +1$ is even, so that
\[
\max_{i\in \check{\beta}_{\kappa}} d_{i,\kappa} =\zeta_{\kappa}+\lfloor(T+1)/2\rfloor -\lfloor (\varphi_{\kappa}+1)/2\rfloor \geq t_{\kappa}
\]
If $\varphi_{\kappa}$ is even, then $\varphi_{\kappa} =  \lfloor (\varphi_{\kappa}+1)/2\rfloor  +  \lfloor \varphi_{\kappa}/2\rfloor$ and $\varphi_{\kappa} +1$ is odd, so that
\begin{align*}
\max_{i\in \check{\beta}_{\kappa}} d_{i,\kappa} & \geq \zeta_{\kappa}+\lfloor(T+1)/2\rfloor -\lfloor (\varphi_{\kappa}+1)/2\rfloor - 1 \\
& \geq \varphi_{\kappa} - \big( \lfloor (\varphi_{\kappa}+1)/2\rfloor + 1 \big) + \lfloor(T+1)/2\rfloor \geq t_{\kappa}
\end{align*}
Now, starting from any node in $\check{\beta}_{\kappa}$, $G(M_{\kappa})$ is observed and connected after at most $T$ waves, so that $\dot{\beta}_{\kappa}$ can be identified. Whereas starting from any node in $\dot{\beta}_{\kappa}$, all the nodes $\check{\beta}_{\kappa}$ are identified after $T$ waves by virtue of RIOP because $d_{i,\kappa} \leq T$ for any $i\in \check{\beta}_{\kappa}$. 
\end{proof}

\noindent Lemma \ref{lemma:simpleextension}.
\begin{proof} \label{proof-4}
The motif $\kappa$ is observed starting from any node in $\ddot{\beta}_{\kappa} = \{ i\notin \beta_{\kappa}^{\bullet}:\lambda_{i,\kappa}\leq T/2 \}$ by at most $t = T/2+1$ waves, where $t\leq T$ if $T\geq 2$. Thus, $\beta_{\kappa}^* \subseteq \beta_{\kappa}$. Given $\beta_{\kappa}^{\bullet}$ is a restricted set, it remains to show that all the nodes $\beta_{\kappa}^*$ can be identified as $T$SBS ancestors starting from any node in $\ddot{\beta}_{\kappa}^{\bullet}$, and vice versa. Starting from any node in $\ddot{\beta}_{\kappa}$, any other node in $\ddot{\beta}_{\kappa}$ can be identified after at most $2t-2=T$ waves, since at most $t-1$ waves are needed after observing $\kappa$, as well as any node in $\beta_{\kappa}^{\bullet}$ after at most $t-1+\lfloor (T+1)/2 \rfloor-1=T-1$ waves by Lemma \ref{lemma:combineddefault}. Vice versa by virtue of the RIOP.
\end{proof}

\noindent Lemma \ref{lemma:betas}.
\begin{proof} \label{proof-5}
On the one hand, a sample node $i$ with $d_{i,\kappa}(G_s) \leq T$ must be a $T$SBS ancestor of $\kappa$ in $G$, since $d_{i,\kappa}(G_s) \geq d_{i,\kappa}(G)$ generally. On the other hand, even when a sample node $i$ with $d_{i,\kappa}(G_s) > T$ is a $T$SBS ancestor in $G$, it cannot be verified in the realised $G_s$. 
\end{proof}

\bibliographystyle{agsm} % Harvard style
\bibliography{Bibliography_TSBS}

\end{document}